\documentstyle[preprint,aps,floats]{revtex}
\tighten

 \input amssym.def   
 \input amssym.tex

\input BoxedEPS
\SetRokickiEPSFSpecial  %% for dvips by Tom Rokicki, for VMS
\HideDisplacementBoxes

\font\twelvemsb=msbm10 at 12pt
\font\ninemsb=msbm7 at 9pt
\font\sixmsb=msbm5 at 6pt
\newfam\Msbfam
\textfont\Msbfam=\twelvemsb
\scriptfont\Msbfam=\ninemsb
\scriptscriptfont\Msbfam=\sixmsb

\def\half{{\textstyle{1\over2}}}
\def\quar{{\textstyle{1\over4}}}

\def\beq{\begin{equation}}
\def\eeq{\end{equation}}
\def\bi{\begin{itemize}}
\def\ei{\end{itemize}}
\def\beqar{\begin{eqnarray}}
\def\eeqar{\end{eqnarray}}

\newcommand{\Ee}{\mbox{$\cal E\;$}}

\newcommand{\Ll}{\mbox{$\cal L\;$}}

\newcommand{\r}{{\bf r}}  %%% Note: \mbox{\bf .. not good here! - Marty
\newcommand{\k}{{\bf k}}

\newcommand{\A}{{\bf A}}
\newcommand{\B}{{\bf B}}
\newcommand{\E}{{\bf E}}

\newcommand{\bp}{{\bf p}}
\newcommand{\bj}{{\bf j}}

\newcommand{\rmd}{{\rm d\null}}

\def\boldnab{\mbox{\boldmath$\nabla$}}

\newcommand{\mn}{{\mu\nu}}
\newcommand{\nm}{{\nu\mu}}
\newcommand{\ab}{{\alpha\beta}}

\newcommand{\nin}{\noindent}

\let\varkappa\kappa

\skip\footins = 14pt % was 12
\begin{document}

%% \nonfrenchspacing
%% \flushbottom

\title{Chern-Simons Violation of Lorentz and PCT Symmetries \\ in
Electrodynamics}

\author{R. Jackiw\footnotemark[1]}

\footnotetext[1] {\baselineskip=12pt This work is supported in part by funds
provided by  the U.S.~Department of Energy (D.O.E.)\\ under contract
\#DE-FC02-94ER40818.  \quad MIT-CTP-2796\quad
hep-ph/9811322\quad  November 1998\\ Workshop on Lorentz violation,
Bloomington, Indiana, November 1998\\ Orbis Scientia 1998, Coral Gables, Florida,
December 1998}

\address{Center for Theoretical Physics\\ Massachusetts Institute of
Technology\\ Cambridge, MA ~02139--4307}

\maketitle
\begin{abstract}\noindent
Recently proposed Lorentz-PCT noninvariant modifications of electromagnetism
are reviewed. Their experimental consequences are described, and it is argued that
available data decisively rules out their occurrence in Nature.
\end{abstract}

%\setcounter{page}{0}
%\thispagestyle{empty}

%\newpage

\bigskip

\noindent
The principle of special relativity is very firmly established in the minds of
physicists, and it is experimentally confirmed, without known exception. 
Nevertheless, today's availability of high-precision instruments
%% , capable of precise
%% measurement, 
lets us ask whether this principle is only approximately true, and
leads us to seek possible mechanism for its violation.  Such an inquiry is not
unreasonable, since we know that a relativity principle does {\bf not} apply to the
discrete transformations of time and space reversal.

Special relativity arose when the symmetry of Maxwell's electrodynamical field
theory, i.e., Lorentz invariance, was elevated to encompass particle mechanics,
whose Newtonian, Lorentz noninvariant dynamics had consequently to be
modified.  Therefore violation of special relativity can be looked for in particle
mechanics, in electromagnetism, or in both.  I shall restrict my attention to possible
nonrelativistic behavior in electromagnetism.  

Let me record the conventional equations, both in compact Lorentz covariant, and
in explicit vectorial notation.  We are dealing with the electric and magnetic fields
$(\E, \B)$, that are components of a second-rank antisymmetric tensor $F_\mn
=-F_\nm$, or of its dual ${}^* \! F^\mn = \half
\epsilon^{\mn\ab} F_{\ab}$:
\begin{eqnarray*}
F_{oi} &=& \half \epsilon^{ijk} \, {}^* \! F^{jk} = E^i \\
 \half \epsilon^{ijk} F_{jk}  &=& {}^* \! F^{oi} = -B^i\ .
\end{eqnarray*}
They satisfy the homogeneous Maxwell equations
$$
\boldnab \times \E + \frac{1}{c}\, \frac{\partial \B}{\partial t} = 0 \ , \quad 
\boldnab \cdot \B = 0 
$$
or 
$$
\partial_\mu  \, {}^* \! F^\mn = 0\ .
$$
which permit writing the fields in terms of the potentials $A^\mu = (\phi, \A)$, by
formulas that are invariant against gauge transforming the potentials:
$$
\phi \to \phi - \frac{1}{c}\, \frac{\partial}{\partial t}  \alpha \ , \quad
\A \to \A + \boldnab \alpha
$$
$$
A_\mu \to \A_\mu - \partial_\mu \alpha
$$
These formulas are
$$
\E = -\boldnab \phi  - \frac{1}{c}\, \frac{\partial \A}{\partial t}\ , \quad
\B = \boldnab \times \A
$$
$$
F_\mn = \partial_\mu A_\nu - \partial_\nu A_\mu\ .
$$
The second set of Maxwell's equations, which sees the sources of charge density
$\rho$ and current density $\bj$, $j^\mu = (c\rho, \bj)$, reads
$$
\boldnab \times \B - \frac{1}{c}\, \frac{\partial \E}{\partial t} = \frac{4\pi}{c} \bj
\ , \quad \boldnab \cdot \E = 4\pi\rho
$$
or 
$$
\partial_\mu F^\mn =  \frac{4\pi}{c}  j^\nu
$$
and can be derived from the Lagrange density
$$
\Ll = \frac{1}{8\pi} (\E^2 - \B^2) - \rho\phi + \frac{1}{c} \bj \cdot \A = -
\frac{1}{16\pi} F_\mn F^\mn - \frac{1}{c} j_\mu A^\mu
$$
where the basic variables are the potentials, and the electromagnetic fields are
expressed in terms of them.  Consistency of the equations of motion and gauge
invariance of the Lagrangian formalism require that the charge density and
current satisfy a continuity equation
$$
\frac{\partial}{\partial t} \rho + \boldnab \cdot \bj = 0 = \partial_\mu j^\mu\ . 
$$

Let us now turn to modifications.  In the most obvious departure from the
standard formulas, we add a ``photon mass term'' by supplementing $\Ll$ with
$\frac{\mu^2}{2} A^\mu A_\mu = \frac{\mu^2}{2} \phi^2- \frac{\mu^2}{2}\A^2$
where
$\mu$ has dimension of inverse length.  
In the new equations of motion $-\mu^2 A^\mu$ is added to $j^\mu$, so that when
the wave {\it Ansatz\/} $e^{ik_\alpha x^\alpha} = e^{i(\omega t - {\bf k} \cdot {\bf
r})}$, $k^\alpha = (\omega/c, \k )$, $ k \equiv |\k|$ is taken for fields  in the
source-free case ($j^\mu = 0$), the dispersion law reads
$$
k^\alpha k_\alpha = \mu^2\ ,\qquad \omega = c\sqrt{k^2 + \mu^2}\ .
$$
 Of course, this does not violate Lorentz
invariance -- the mathematical expression of the special relativity principle --
but it destroys  Einstein's physical reasoning that led him to special relativity: light
no longer travels with a universal velocity in all reference frames, and ``$c$''
becomes a mysterious limiting velocity that is not attained by any physical
particle.  Gauge invariance appears to be violated, but today we know that
the gauge principle can be obscured by subtle symmetry-breaking mechanisms,
for example, the mass $\mu$ could arise from a feeble Higgs effect. 
%% Experimental limits on this modification come from geomagnetic data
After solving the modified field equations with prescribed sources, one finds
electromagnetic fields that are distorted by the mass term. Comparison to
experiment is made with geomagnetic data, leading to the limit
$\mu<3\times 10^{-24}$ GeV [A. Goldhaber and M. Nieto, Rev.\ Mod.\ Phys.\ {\bf
43}, 277 (1971)] while observations of the galactic magnetic field give
$\mu<3\times 10^{-36}$ GeV [G. Chibisov, Usp.\ Fiz.\ Nauk {\bf 119}, 551 (1976);
Sov.\ Phys.\ Usp.\ {\bf 19}, 624 (1976)] 
(1 GeV $\sim 10^{13} {\rm cm}^{-1}$).

Lorentz invariance and the relativity principle, but not rotational invariance,
disappear if the Lagrange density is modified by the addition of a further
$\frac{1}{8\pi} \B^2$ term, proportional to $\epsilon$.  However, by rescaling $\A$,
one sees that this is equivalent to redefining the velocity of light from $c$ to
$c_\epsilon \ne c$, in the electromagnetic part of the theory while retaining $c$ as
a parameter in the (unspecified) matter kinematics.  Hence this modification can be
exported into the matter sector and I shall not discuss it further.  It has been
studied by S. Coleman and S. Glashow [Phys.\ Lett.\ B {\bf 405}, 249 (1997)], who
use cosmic ray data to bound the magnitude of the addition by $10^{-23}$.

I now come to yet another modification, introduced by S. Carroll, G. Field, and me
almost a decade ago, which  recently came again to attention  [Phys.\ Rev.\ D
{\bf 41}, 1231 (1990)].  To begin, let us note that in addition to $-\half F_\mn
F^\mn =
\E^2 -
\B^2$, another Lorentz scalar, quadratic in the field strengths, can be constructed:
$-\quar
\,{}^*\!F^\mn F_\mn = \E
\cdot \B$.  However, adding this to the electromagnetic Lagrange density does not
affect the equations of motion, because that quantity, when expressed in terms of
potentials -- the dynamical variables in a Lagrangian formulation -- involves
total derivatives, which do not contribute to equations of motion:
\begin{eqnarray*}
\half \, {}^* \! F^\mn F_\mn &=& \half  \partial_\mu (\epsilon^{\mu \alpha \beta
\gamma} A_\alpha F_{\beta\gamma}) \\
-2 \E \cdot \B &=& \frac{\partial}{\partial t} \Big(\frac{1}{c} \A \cdot \B) + 
\boldnab \cdot (\phi \B - \A \times \E) \ . 
\end{eqnarray*}
However, when the $\E \cdot \B$ quantity is multiplied by another
space-time--dependent field
$\theta(t, \r)$, the total derivative feature disappears and such an addition would
affect dynamics.  Once again using the freedom to modify a Lagrange density by
total derivatives, we see that the addition of $\theta \, {}^* \! FF$ to the Lagrange
density is equivalent to adding $-\partial_\mu \theta \epsilon^{\mu \alpha \beta
\gamma}  A_\alpha F_{\beta\gamma}$.  If $\theta$ is a dynamical field, then the
extended electromagnetism $+~\theta$ system remains Lorentz invariant. 
We shall however posit that neither $\theta$ nor $\partial_\mu \theta$ are
dynamical quantities; rather $\partial^\mu \theta$ is a constant 4-vector $p^\mu =
 (m, \bp)$ that picks out a direction in space-time, thereby violating Lorentz
invariance.  Thus we are led to consider an electromagnetic theory, where the
conventional Maxwell Lagrange density is modified by
\begin{eqnarray*}
\Delta \Ll &=&- \quar p_\mu (\epsilon^{\mu \alpha \beta
\gamma} A_\alpha F_{\beta\gamma}) = -\half p_\mu  \, {}^* \! F^\mn A_\nu \\
 &=& -\half m \A \cdot \B + 
\half \bp \cdot (\phi \B - \A \times \E) \ .
\end{eqnarray*}
The sourceless Maxwell equations are unchanged (the fields continue to be
expressed by potentials).  Only the equations with sources are changed, and the
change can be viewed as a field-dependent addition to the source current.
\begin{eqnarray*}
\partial_\mu F^\mn &=& \frac{4\pi}{c} j^\nu + p_\mu \, {}^* \! F^\mn \\
 \nabla \times\B - \frac{1}{c} \frac{\partial\E}{\partial t}  = \frac{4\pi}{c} \bj
-m \B &+& \bp \times \E \ , \quad \boldnab \cdot \E = 4\pi \rho - \bp \cdot \B\ .
\end{eqnarray*}
Note that the field equations are gauge invariant, even though the Lagrange
density is not.  The quantities $m$ and $\bp$ have dimension of inverse length;
the latter breaks rotational invariance by selecting a direction in space; $m$
breaks the invariance of the theory against Lorentz boosts.  Presumably for the
interesting case we should select vanishing $\bp$ (rest frame of $p^\mu$, which is
taken to be time-like) so that rotational isotropy is maintained.  Parity is also
broken, since the pseudoscalar
$\B$ mixes with vector $\E$, but 
charge conjugation and time inversion remain intact, hence PCT is broken.

Before describing the consequences of our model, let me recall some facts about
the various quantities that we have introduced. ${}^* \! F^\mn F_\mn$ is the
so-called Chern-Pontryagin density; its non-Abelian generalization plays an
important role in the ``standard'' particle physics model, where it is a measure of
the anomalous (quantum mechanical) nonconservation of the axial vector
current.  It is responsible for the decay of the neutral pion to two photons, and for
proton decay.  The 4-vector whose divergence gives ${}^* \! F^\mn F_\mn$  is
called the Chern-Simons density.  Both objects are templates for the topologically
nontrivial behavior of non-Abelian gauge fields.  

In some extensions of the standard model, ${}^* \! F^\mn F_\mn$  is coupled to a
further dynamical field, like the $\theta$-field mentioned above, which describes
a hypothetical particle -- the axion -- whose role is to ensure CP symmetry of
strong interactions.  However, no evidence for such a particle has been found thus
far.

Note further that if we were living in $(2+1)$-dimensional space-time, that is on a
plane rather than in a three-dimensional volume, the $\epsilon$-tensor would
have only three indices and we could introduce the Chern-Simons term into
$(2+1)$-dimensional electrodynamics without the external 4-vector $p^\mu$, i.e.,
$(2+1)$-dimensional Lorentz invariance would be preserved by $\Delta \Ll = m
\epsilon^{\alpha\beta\gamma} A_\alpha F_{\beta \gamma}$.  
Chern-Simons modified electrodynamics plays a role in planar electromagnetic
phenomena, as in the quantum Hall effect, and perhaps also in high-$T_c$
superconductivity.

Finally we remark that with vanishing $\bp$ and absence of sources, so that $\E =
0$, the remaining modified Maxwell equations read $\boldnab \times \B = -m\B$,
$\boldnab \cdot \B = 0$.  These have arisen previously in
magnetohydrodynamics.  They coincide with the conventional Maxwell equations
in the presence of neutral sources and steady currents $(\rho=0, \boldnab \cdot
\bj =0)$ and are seen to be equivalent to the conventional Amp\`ere's law, when
the further condition is imposed that $\bj$ is proportional to $\B$.

What is the consequence of our Lorentz invariance violating modification?

Let us examine wave solutions in the absence of sources.  $(\rho = \bj = 0, j^\mu
=0)$.  We again make the {\it Ansatz\/} that fields behave as exponentials of 
phases, 
$e^{i(\omega t - {\bf k} \cdot {\bf r})} = e^{ik_\alpha x^\alpha}$, $ k^\alpha =
(\omega/c, \k )$, $ k \equiv |\k|$,
and find the dispersion law
$$
(k^\alpha k_\alpha)^2 + (k^a k_\alpha) (p^\beta p_\beta) = (k^a p_a)^2\ .
$$
From this one can show that introducing $p^\alpha$ has the consequence of
splitting the photons into two polarization modes, each traveling with different
velocities $\omega/k$ -- forceful evidence of Lorentz and parity violation.  This is
very easily seen in the rotation invariant case, $\bp = 0$.  One finds
$$
\omega^2 = ck (ck \pm mc)\ .
$$
Note that $\omega$ can become imaginary for modes with $k<m$.  This means
there are unstable runaway solutions.  These do not contradict energy
conservation, because the energy $\Ee$ is no longer the positive expression
of the Maxwell theory, $\half \int \rmd^3 r (\E^2 + \B^2)$.  Rather we now
have
$$
\Ee = \half \int \rmd^3 r \Big[ \E^2 + \Big(\B + \frac{m}{2} \A\Big)^2 \Big]
- \frac{m^2}{8} \int \rmd^3 r \A^2 \ .
$$
With unstable solutions each of the two terms contributing to $\Ee$ grows without
bound, yet $\Ee$ stays finite and time-independent owing to a cancellation
between the two.  (The energy is gauge invariant -- in spite of appearances.) 
However, runaway, exponentially growing modes can be avoided by allowing for
noncausal propagation for well-behaved sources (similar to the way runaway
solutions are eliminated from the Abraham-Lorentz equation of conventional
electrodynamics).

Returning now to our plane wave solutions, we observe that a plane-polarized
wave -- a superposition to two circularly polarized modes traveling at different
velocities -- will be rotated when it travels through space.  Since $p^\alpha$ is
small, we can solve for $\omega$ to first order in $p^\alpha$, and find, (without
setting $\bp$ to zero)
$$
k=\frac{\omega}{c} \mp \half (m- \bp \cdot \hat{\k})
$$
so the change $\Delta$ in the polarization, as the wave travels a distance $L$ is 
$$
\Delta = -\half (m- \bp \cdot \hat{\k}) L\ .
$$
This is similar to the Faraday effect, where a polarization change is induced by
ambient magnetic fields.  However our phase change is wavelength independent,
while the Faraday effect rotation is proportional to wavelength squared.  So the
two effects can be distinguished.

When comparing predictions of this theory to experimental data, Carroll, Field and
I assumed that rotation invariance holds, we set $\bp = 0$, and the entire effect is
parameterized by $m$ (time-like $p^\alpha$).  Geomagnetic data can be
confronted with the distorted magnetic field that solves the modified equations in
the presence of sources. But the data is somewhat difficult to interpret in our
context, and the most plausible limit is 
$$
m \le 6 \times 10^{-26} {\rm GeV}\ .
$$
However, examining the polarization of light from distant galaxies and removing
the rotation due to the Faraday effect,  yields a much more stringent result
$$
m \le  10^{-42} {\rm GeV}
$$
[see also M. Goldhaber and V. Trimble, J.~Astrophys.~Astr.\ {\bf 17}, 17(1996)].
Since effects of nonzero $m$ can appear only at distances greater than the
associated Compton wavelength, which for the above is the distance to the horizon,
astrophysical data apparently rules out nonvanishing $m$.

How about vanishing $m$ and nonvanishing $\bp$ (space-like $p^\mu$)?

Our formula for polarization change indicates that here too there should be a
non-Faraday rotation.  Moreover, one can show that space like $p^\alpha$ produce
no instability.  However,  Carroll, Field, and I believed that such a violation
of rotational symmetry is unlikely.

Thus, we were very surprised when there appeared a Physical Review Letter by B.
Nodland and J. Ralston [Phys.\ Rev.\ Lett.\ {\bf 78}, 3043 (1997)] 
alleging that precisely this kind of anisotropy exists. Evidence for this startling
assertion was drawn from the same galactic data that we used in our analysis,
which gave us the null result.

We were not the only ones surprised.  Here is a sampling of news stories about this
``discovery'' in the popular and semipopular press, following a report of the
Nodland-Ralston ``result'' issued by the American Institute of Physics.
\bigskip

\nin American Institute of Physics

``Is the universe birefringent?'', 17 April 1997

\nin {\it New York Times}

``\thinspace `This (don't ask which) side up' may apply to the universe'', 18 April
1997

``Theory about the universe has its ups and downs'', 25 April 1997

\noindent Associated Press

``Space isn't the same in all directions'',  18 April 1997

\nin \textit{Time}

``This side up.  New evidence challenges Einstein's universe'',  18 April 1997

\nin \textit{Tabloid}

``Einstein was wrong'', 19 April 1997

\nin \textit{Science News}

``Does the cosmos have a direction?'',  26 April 1997

``Cosmic axis begets cosmic controversy'', 10 May 1997

\nin \textit{Sky and Telescope}

``No lopsided universe'',   9 May 1997

\nin \textit{Galileo} (Italy)

News report,   15 May 1997

\nin \textit{Physics World} (UK)

``Axis of universe debate rumbles on'',   15 May 1997

\nin \textit{Los Angeles Times}

``A zigzag route to the truth'',    7 August 1997

\nin \textit{Popular Science}

``Which way is up?'',     October 1997

\nin \textit{ABQ Journal}

``VLA data defends big bang theory'',     17 November 1997

\nin \textit{Scientific American}

``Twist and shout'',  (web page) 
\bigskip

Interest in the result also evoked  humorous reactions, in the form of a syndicated
cartoon by Hilary Price depicting existential anguish engendered by life in an
anisotropic universe, a statement by Lyndon La Rouche that he knew it all the time
(interview, 7 May 1997 with A.~Papert), and a claim for extraterrestrial life
(http://www.enterprisemission.com).

\begin{figure}[ht]
\centerline{\BoxedEPSF{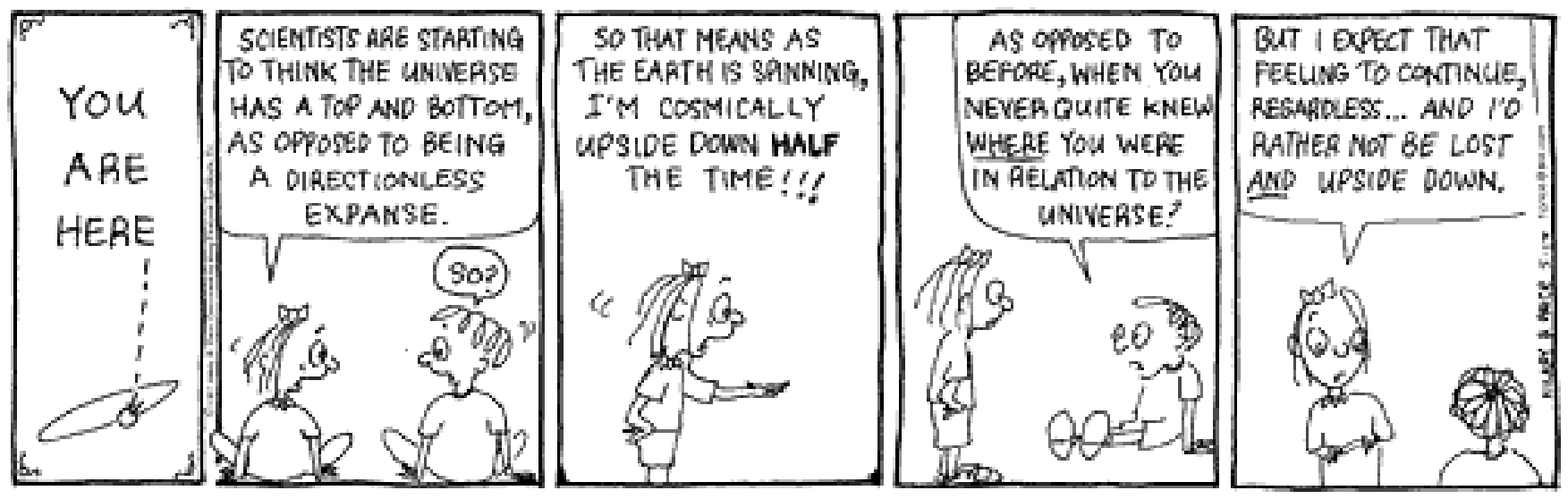 scaled 900}}
\centerline{Hilary Price \copyright King Features Syndicate Inc.\ 1997}
\end{figure}

Unfortunately, it appears that Nodland and Ralston made a mistake in their data
analysis.  Carroll and Field [Phys.\ Rev.\ Lett.\ {\bf 79}, 234 (1997)] reanalyzed
the data, identified their error, and found no anisotropy.  Thus our original
conclusion that there is no evidence for a Chern-Simons modification to
electromagnetism stands, and has been confirmed by several other investigations. 
The entire matter is nicely reviewed on
http://ITP.UCSB.edu/$\sim$carroll/aniso.html\ .

In spite of the negative results, we can nevertheless draw an interesting
conclusion.  We know that in Nature parity P, charge conjugation C,
and time reversal T are violated.  While local field theory and Lorentz invariance
guarantee that PCT will be conserved, it remains an experimental and
interesting question whether PCT is valid in Nature, which perhaps does not
make use only of local field theoretic dynamics.  But PCT violation in any
corner of a grand unified theoretical structure would induce PCT violation in
electrodynamics, so the stringent limits that we put on the PCT-violating
Chern-Simons term, also limit some forms of PCT violation anywhere else in the
``final theory''.

\end{document}